\documentstyle[11pt,paspconf,epsf]{article}

\def\etal{{\it et al.}}

\begin{document}
\title {TESTING ENVIRONMENTAL INFLUENCES ON STAR
 FORMATION WITH A SAMPLE OF LSB DWARF GALAXIES
 IN THE VIRGO CLUSTER}
\author{Ana Heller, Elchanan Almoznino  \& Noah  Brosch}
\affil{The Wise Observatory and 
the School of Physics and Astronomy \\ Tel Aviv University, Tel Aviv 69978,  
Israel}

\begin{abstract}
  We analyze   the star formation activity of  an homogeneous
 LSB dwarf galaxy sample
  in the Virgo cluster, as a function of the    radial velocity  relative
 to the cluster mean velocity and the projected distance from  the center of the cluster,
 using CCD images obtained at the
   Wise Observatory.
 The localized H$\alpha$ emission in the HII regions of this sample
 is compared  to that of an isolated gas-rich sample of LSB dwarf galaxies  and 
that of  a representative sample of Blue Compact Dwarf (BCD) galaxies in the cluster. 
 We report preliminary results on the LSB dwarf star formation histories 
obtained from  surface color distribution.  
 \end{abstract}
%\vspace{0.5cm}

%\vspace{1.0 truecm}
\section{Introduction}
 There is  an endless debate in the literature regarding the 
  influence of the neighborhood over the
 physical processes that govern the evolution 
  and   the relative abundance of types in dwarf galaxies (Koopmann 1997).
  Futhermore, recent investigations of distant galaxies  suggest  
 that  while rich cluster galaxies have reduced star formation compared
to field galaxies  of  the same central concentration of light,   the star formation
 rate (SFR) is quite sensitive 
to local  galaxy density, both inside and outside of clusters, but  
 the highest levels of star formation are encountered
 in the intermediate density environment~(Hashimoto {\etal} 1997).
Therefore, we must consider not only the surface brightness of galaxies in each class, but
 also their
environments, to see if they influence the star formation  properties.

The high 
 Galactic latitude of the Virgo cluster (VC), and therefore very small foreground extinction,
 coupled with the relative nearness of the cluster, allows one to study 
 the effect of the environment on  the star formation histories
 of  low surface brightness (LSB)~dwarf galaxies.  Their low mean density and gravitational binding energies (Bothun {\etal} 1985)
  make them 
 susceptible  to dynamical processes that
potentially operate in  the cluster, which has 
a  mean galaxy  density $\sim$10~galaxies Mpc$^{-3}$.
 However, the VC   is not a relaxed, virialized cluster; it shows a complex
 structure  in which the LSB dwarf population has no
 preferential   velocity  distribution.
 A kinematic structure map,  showing the position on
 the sky of the various groups and clouds within the VC cluster area, 
 has been   published by Hoffman {\it et al.}~ (1989). 

External forces produced by 
  tidal interactions or tidal shocks  
 should   depend strongly on the mass of the perturber/perturbed  galaxy, the relative speed, 
 and the minimal distance
between them.  
Other
 effects of the surroundings, 
 such as ram pressure stripping, evaporation, or turbulent viscous stripping,
 are  believed to be  active in gas removal,
with  estimated stripping time-scale $\sim$ 10$^{9}$~yr, and should be
 most efficient 
 in the hot and dense central part of the VC,
 within $\sim$300kpc of M87 ( Ferguson \& Binggeli ~1994). 
 The removal of gas by the cluster environment is also believed to be responsible
 for the dependence of  chemical abundance  properties of HII regions
 on cluster location in late-type spiral galaxies in   VC~(Skillman 
 {\etal} 1995).
 Note that none of the LSB dwarf galaxies included in our sample (see below) 
 is located in the central region of the VC, and all of them are more distant 
 than $\sim$2.5 degrees ( $\sim$1Mpc, projected) from M87.  Therefore,
 the processes mentioned before may not be   affecting these galaxies.
This  is supported by VLA HI maps of the  more  intense sources
 of our sample of galaxies
  (Skillman \& Bothun 1986,  Skillman {\etal} 1987)) and   Arecibo
 HI maps (Hoffman {\etal} 1996,  Salpeter \& Hoffman 1996). We found that  all  mapped galaxies support 
the previous finding of Skillman {\etal} (1987);   at specific projected distances from the
cluster center (R$_{M87}$),
 the derived   HI   to  optical   diameter ratios 
 (D$_{HI}$/D$_{opt}$)  are  larger than   those of a sample of
 VC spiral galaxies. 

 However,  long before passing near  the cluster core region,
the pressure of the intracluster medium (ICM)
may induce  star formation in a gas-rich galaxy,
  due to processes such as
  compression of gas clouds, density enhancements, accumulation 
of gas in clouds that later  collapse gravitationally, or cloud-cloud collisions 
in  the interstellar medium (ISM)~(Elmegreen~1997). Pressure confinement
may also prevent or reduce the outflow of gas driven by supernovae or 
winds from OB  stars ~(Babul \& Rees~1992).  
If these  dynamical effects 
  are capable  to significantly drive  the star population and evolution of
 LSB dwarf galaxies in the VC,     we should then expect 
  to see systematic differences
  in their  SFR related to the location within the cluster.
  
\section{The sample}
In the  Virgo Cluster Catalog~(VCC, ~Binggeli, Sandage \& Tammann~1985),
  the surface brightness  serves  
 as  a   luminosity class indicator for late-type galaxies.
 The highest  surface brightness objects  are  assigned to class~III 
 while those with the lowest surface brightness belong to  class~V. 
There are 31 galaxies in the VCC  with certain classification 
 ImIV and ImV; these are LSB galaxies
with mean surface brightness  fainter than 
  24-25 mag arcsec$^{-2}$. We rejected two galaxies
 from the 24 ImIV to ImV galaxies with  non-zero
 HI measurements ~(Hoffman {\etal} ~1987):   
     a  small, faint 17.5~magnitude galaxy which fell below
  our threshold,  and another
  with  very high heliocentric  radial velocity, which violated our  
  membership VC  criterion  of  $v_{ \odot} <  3000$ km sec$^{-1}$.
 This $v_{ \odot}$ restriction arises because of 
the   void  behind the VC, between the W and M clouds, from
  2800 km/sec
   to 3500 km/sec, where  no galaxies
 are detected~(Binggeli {\etal} ~1993).
Here we consider all members in the various clouds
as part of the  VC.  The mean heliocentric 
 radial velocities of the galaxies is 1200 km/sec, close to the  mean velocity
  of  dE galaxies,  1139$\pm$67 km/sec,
  though   slightly
higher than the  mean velocity of the cluster 
(1050$\pm$35km/sec; Binggeli {\etal} ~1993).

 Our final  sample   sample  comprises of     27 galaxies; it
  includes  22 galaxies  of type  ImIV  to ImV, 
  four of uncertain classification  ImV/dE, ImV? or Im:,  and one    Im III-IVpec.
The limiting  magnitude for inclusion in our sample
  is $ m_B=17.5$; the galaxies are small and their  major-axes
  range  from  16  to  120~arcsec at
  the 25 ~mag~arcsec$^{-2}$  isophote.

\section{Observations, Results and Analysis} 
In order to study the  on-going
star formation we observed the galaxies with  a  narrow\--band filter
 in the red continuum near H$\alpha$ (H$\alpha$-off) and with 
a set of narrow-band filters centered on the rest-frame H$\alpha$ ~line (H$\alpha$-on)
 of the galaxies. Deep images in the broad-band filters  U, B,V, R and~I 
 provided  some  constraints on the older stellar populations. 
 The narrow-band  H$\alpha$ images were taken during  
the observing runs of 1996
 and 1997 at the Wise Observatory~(WO).  Two of the faintest galaxies  were
 observed with   narrow-band H$\alpha$  filters  at the SAO 6m telescope. 
 The H$\alpha$-on  images were calibrated with observations of
 spectrophotometric standards,
 co-added, 
and  the H$\alpha$-off image was subtracted  from each one to produce  final net-H$\alpha$ images.   The  H$\alpha$ images  of the sampled galaxies
 are shown in  Heller {\it et al.}~ (1998). The typical limiting H$\alpha$ flux is 
10$^{-16}$ erg~sec$^{-1}$~cm$^{-2}$;
the limiting  spatial resolution is ~300-400 pc.  
All UBVRI images were collected at the WO
 during  1997 and 1998 and   standard stars  (Landolt 1973, 1992)
  were used for  calibration.

We detected H$\alpha$ emission in
62\%   of  the  sample galaxies. The detection rate  is higher
when considering only those galaxies with certain  ImIV to ImV classification
 (68\%).
Three of the galaxies in which we did not detect H$\alpha$ emission  were those 
with uncertain classification (dE2 or ImIV, Im?) in the catalog.
 Their HI fluxes  put them in low  HI content  group  
  with  a line flux integral   $\leq$776 mJy Km/sec. 
   The high HI content group, with an equal number of galaxies, had
  a line flux  integral $\geq$950 mJy Km/sec.

We calculated the total  SFR of the galaxies as in  Kennicutt ~{\it et al.}~ (1994),\\ 
SFR=2.93~10$^{11}$~F(H$\alpha$), where F(H$\alpha$) 
is the total line flux  in cgs units and the SFR is 
in M$_{\odot} $yr$^{-1}$.
  We adopted  a common  distance of 18 Mpc for all the galaxies of our sample. This
 may not be exactly true, but allows a comparison  with the values calculated
in the same way for  BCD galaxies in the VC  ( Almoznino~1996).
The typical SFR of our LSB sample is  0.007 M$_{\odot}$yr$^{-1}$ 
a factor 10 lower than BCDs.
The typical  H$\alpha$  equivalent   width (EW) of the LSB galaxies
 is  $\sim$30{\AA}  peaking at 
100{\AA},  a factor 2 lower than the BCDs.

Some of the LSB galaxies in the sample have low heliocentric velocities;
it is possible that these are objects falling into the VC from its distant side
(Tully \& Shaya 1984). 
 We tested the possibility that these galaxies may have enhanced star formation 
 because of interaction with the cluster gas, but this did not prove out.
 Likewise, we did not find  any clear dependence between the net-H$\alpha$ emission,
 the EW,  
 the relative velocity of a galaxy with respect to the mean  cluster velocity,
 the angular distance to the VC center, subclustering, or   the HI flux of a galaxy.  
 
\begin{figure}[tbh]
%\begin{figure}
\vspace{6cm}
%\ref{Virgopred}
%\plotone{hist.cardif.ps}
\includegraphics{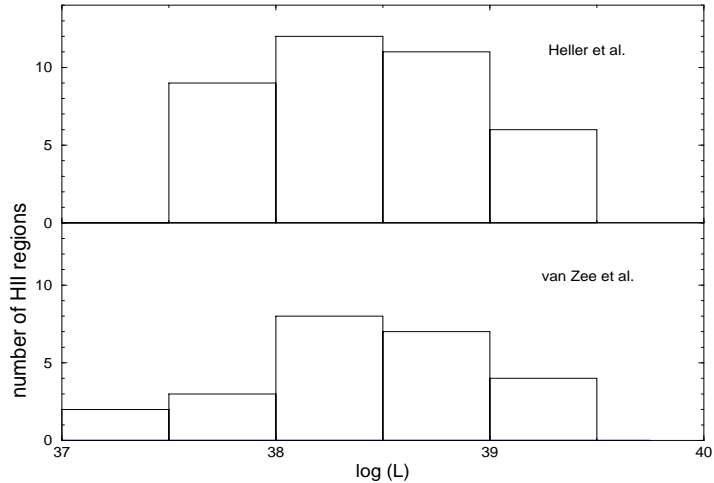}  
\caption{Distribution of the logarithm of the luminosity  for individual HII regions,
  Virgo Cluster LSB dwarf sample (Heller  {\etal} 1998) (up),
 isolated gas rich  LSB  sample (van Zee~1998) (down).} \label{fig-1}
\end{figure}

A number of  tests were performed on the level of individual HII regions. We first
  compared the range of  net-H$\alpha$  fluxes of individual HII 
regions of  our LSB sample 
 to  those of isolated, gas rich  LSB galaxies  studied by  van Zee~(1998) and 
 found that they are extremely  similar (Figure~\ref{fig-1}).
 
 Both samples have approximately the same number of HII regions per galaxy 
 and  in both the luminosity peaks at $\sim$3 10$^{39}$ erg s$^{-1}$.
 The results indicate  no differences between galaxies in  the cluster and  isolated ones as long as they are dwarfs LSBs,
 although it should be noted that
the covering factor of a galaxy by HII regions  is much lower in
 the isolated sample, originally selected to have extended HI
 envelopes. 

\begin{figure}[tbh]
\vspace{10cm}
%\ref{Virgopred}
\includegraphics{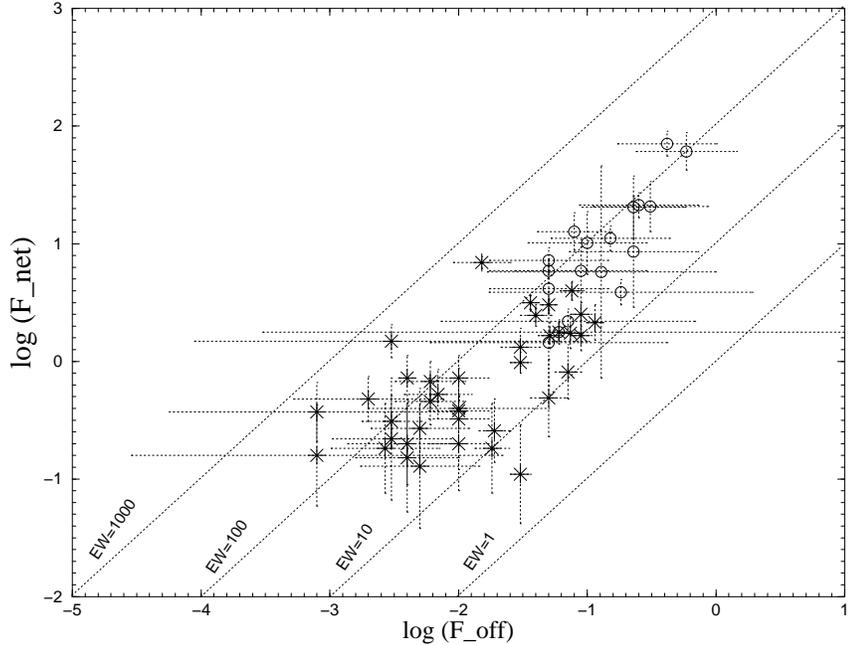}
\caption{Correlation between the logarithms of
 the H$\alpha$ emission 
and of the red continuum light
for individual HII regions in LSB galaxies.  The data for HII regions of
BCDs is plotted as circles and that for the HII regions
of LSBs is represented by stars. The units are 10$^{-14}$ erg cm$^{-2}$ s$^{-1}$
for the line flux and 10$^{-14}$ erg cm$^{-2}$ s$^{-1}$ \AA\,$^{-1}$ for
the continuum flux density. The dotted lines  indicate   EW=1000, 100, 
10  and 1{\AA}  respectively.} \label{fig-2}
\end{figure}

In Figure~\ref{fig-2}  we plotted  the line and continuum fluxes of the HII regions 
of each galaxy,  including the HII regions of the BCD sample.
  The dotted lines  indicate   EW=1000, 100, 
10  and 1{\AA}  respectively.  The dearth of HII regions with low EW
  is easy to understand. It is the result of our detection technique,
 where low EW HII regions can hardly be distinguished against the 
 red continuum background. On the other hand, it is not clear what
 limits the high end of the EW distribution.
It is very interesting to note that both samples align 
with approximately the same EW for their HII regions
 between EW=10   and  100{\AA}.

\begin{figure}[tbh]
\vspace{10cm}
%\plotone {HII.mos3.ps}
%\ref{Virgopred}
\includegraphics{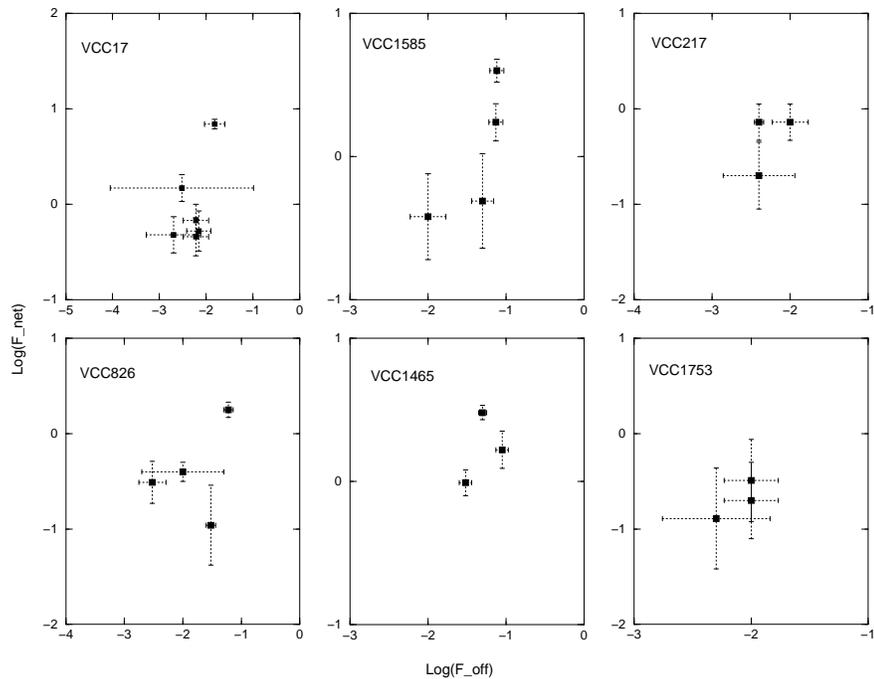}
\caption{Correlation between the logarithms of the H$\alpha$ emission 
and of the red continuum light
for individual HII regions in LSB galaxies with multiple HII regions.} \label{fig-3}

\end{figure}

 In a  model in which the interaction between the ISM and the ICM
 is the dominant star formation trigger, 
 the spin of the galaxy  moves  the HII region from the recent SF burst,
 which is the locus of the interaction with the ICM,
 away from the galaxy leading edge.
 As a burst  evolves,  the  H$\alpha$  EW should decrease.
This is because of the reduction of the line emission (a few Myrs after
the disappearance of the ionizing flux) at the same time 
 as the continuum  increases  due to the net increase in the number
of low mass stars.  Note that this is true for a star burst which takes place
in a pre-existing old population. 
Our results for individual HII regions in each galaxy (Figure~\ref{fig-3})
 do not support this scenario.

An analysis of the number distribution of the star formation regions for these
galaxies (and in other star-forming dwarfs),
  based on the visual inspection of
 the  net-H$\alpha$ images,  results in asymmetric indices ($AI$,  ratio  of the
number of HII regions counted in the poor to the rich area)
$ AI<0.5$  (Brosch {\etal}  1998a and these proceedings).  In other words,
 the  HII regions are    located predominantly on one side of the galaxy,
but are not the result of a bow-shock induced SF. 

\begin{figure}[tbh]
\vspace{10cm}
%\plotone {HII.dv.ps}
%\ref{Virgopred}
\includegraphics{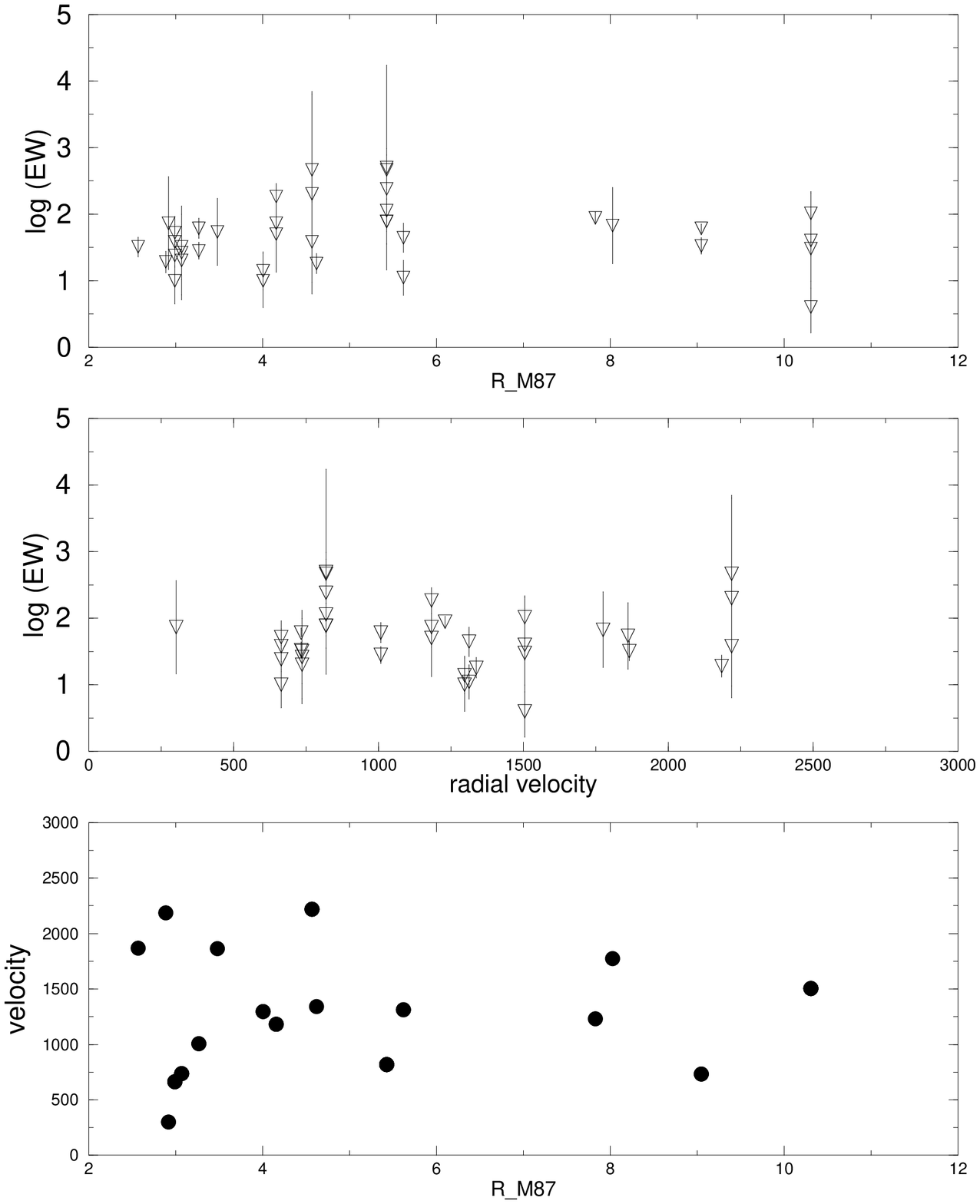}
\caption{Log(EW) vs. the projected angular distance in degrees
 from cluster center (R$_{M87}$) (up).
  Log(EW) vs. heliocentric radial velocity in km/sec (v$_{\odot}$) (middle).
  Heliocentric radial velocity (v$_{\odot}$) vs. R$_{M87}$ (down).} \label{fig-4}
\end{figure}

In  case that the bursts are  the result of both  external and internal agents, 
 we should expect to see a
 broad range of EW values  due to the different locations
of the orbiting  galaxies in the cluster.  For most of the objects this is not the case (Figure ~\ref{fig-4}), and 
 the EW of individual HII regions changes very smoothly in a galaxy, with  no correlation
  with the angular distance to the cluster center (M87) nor
 with the heliocentric  radial velocity of the galaxy.

Lacking a clear understanding of external triggers which may be relevant 
 to explain the star formation in Virgo LSB dwarfs, we turned to internal
 effects which might influence this process. We searched for correlations
 between the SFR and other individual observable parameters of the 
 galaxies in the sample. 
The strongest correlation was found 
 between the  line and continuum  fluxes
of the individual HII regions;
  for more intense line emission the red continuum
 is also more intense.  A linear regression between these variables  
indicates a correlation coefficient of~0.62.
The same calculation for the sample of  17 BCDs in the VC 
 leads to a correlation coefficient of~0.82 (Heller {\etal} 1998).
The result is reminiscent of 
the    correlation found between
the SFR and the mean blue surface brightness of late-type dwarf galaxies
 (Brosch {\etal} 1998b).
We conclude that  the star formation of most  LSB dwarf galaxies in VC depends on internal
processes;  a self regulating  heating-cooling mechanism  modulated by 
the local volume density of  stars  has to be at work, probably  limiting the SFR 
 of  these types of galaxies.

In    order to constraint the star formation histories
we analyzed the surface color distribution    of the galaxies.  
At present we  applied this method to a single object from our sample.
 We   mapped  the color indices   (U-R), (U-B) and (B-V)
  for  VCC826 and 
  we used the  monochromatic H$\alpha$-off  magnitude as R.  We find that 
 the area under the HII region is bluer than the rest of the galaxy. 
 This indicates that there are many young stars under the HII
 region, and it is not just a lop-sided IMF which creates only massive stars.
 Therefore a truncated IMF with only  stars with mass higher 
 than 10 M$_{\odot}$
 cannot  be considered a possible explanation for  the low luminosity  of LSB  galaxies.
The color difference between
the area under  the strong HII region of VCC826  and
 the rest of the galaxy is only $\delta$(U-R) $\sim$ 0.8; this
   indicates that the stellar population of the rest of the galaxy cannot be 
 very  old.  Although we have not yet attempted a full evolutionary model
calculation, possible   combination  which yield this color difference could have typically
 B3V stars under the HII regions and A0V in the rest of the  galaxy,
 or   O6V   under the HII regions     and  B7V in the rest of the  galaxy.
 This is similar to the interpretation of broad-band colors for BCDs 
as due to at least two stellar populations,
 reached by  Almoznino (1996).
 Note that we cannot rule out the existence of  even older stellar generations 
on the basis of the present results.

 \section{Conclusions}
We found no strong evidence for  environmental influences in the SF activity
of LSB dwarf galaxies in the Virgo Cluster.
The SF process seems to be mostly local and   
 regulated by the local population of stars.
The  SF histories of LSB dwarfs  and BCD are probably  similar but the bursts differ in
intensity.
The method of local color analysis proves to be very significant in
constraining possible evolutionary scenarios.

\acknowledgments
AH acknowledges 
support from the
US-Israel Binational Science Foundation and thanks Liese Van Zee for kindly providing
 comparison images.
 EA is supported by a special grant from the Ministry of Science and the 
Arts to develop TAUVEX, a UV space imaging experiment.  NB is grateful for the continued 
support of the Austrian Friends of Tel Aviv University.  Astronomical 
research at Tel Aviv University
is partly supported by a Center of Excellence Award from the Israel Academy of 
Sciences.


\begin{references}

\reference Almoznino, E. 1996, PhD thesis, Tel Aviv University.

\reference Babul, A. \& Rees, M.J. 1992, MNRAS, 255, 346.

\reference Binggeli,  B., Sandage, A. \& Tammann, G.A. 1985, AJ,  90, 1681.

\reference Binggeli,  B., Popescu, C.C. \& Tammann, G.A. 1993, 
 A\&AS,  98, 275.

\reference Bothun, G.D., Mould, J.R , Wirth, A.  \& Caldwell, N. 1985, ApJ, 90, 697.

\reference Brosch, N., Heller, A.B.  \& Almoznino, E. 1998a,  MNRAS, in press.

\reference Brosch, N., Heller, A.B.  \& Almoznino, E. 1998b,  ApJ, 504, 720.


%\reference de Boer, K.S., Braum, J.M., Vallenari, A., \& Melbold, U. 1998, A\&A, 329, L49.

\reference Elmegreen, B.G. 1998, in "Origins of Galaxies, Star, Planets and Life" (C. E.  Woodward,
 H. A. Thronson  \& M. Shull, eds.), ASP series, in press.

\reference Ferguson, H. \& Binggeli, B. 1994,  A\&A Rev, 6, 67.

\reference Hashimoto, Y., Oemler, A., Lin, H.  \&  Tucker, D.L. 1998, ApJ, 499, 589.

\reference Heller, A.B., Almoznino, E. \& Brosch, N. 1988, MNRAS, in press.

\reference  Hoffman, G.L., Helou, G., Salpeter, E.E., Glosson, J.  \& Sandage, A.
  1987, ApJS,  63, 247.

\reference Hoffman, G.L., Helou, G., Salpeter, E.E.  \& Lewis, B. M.
  1989, ApJ,  339, 812.

\reference  Hoffman, G.L., Salpeter, E.E., Farhat, B., Roos, T., Williams, H.  \& Helou, G.
 1996, ApJS, 105, 269.

\reference  Kennicutt, R.C., Tamblyn, P. \& Congdon, C.W. 1994, ApJ, 435, 22.

\reference Koopmann, R.A. 1997, PhD thesis, Yale University.


\reference  Landolt, A.U. 1973, A. J 78, 958.

\reference  Landolt, A.U. 1992, A. J 104, 340.

%\reference Sandage, A.,  Binggeli, B.\& Tammann, G. A.  1985, 
%A J, 90, 9.

\reference Salpeter, E.E.  \& Hoffman, G.L. 1996, ApJ, 465, 5958.

\reference Shields, G. A., Skillman, E.D., Kennicutt, R.C.  \& Zaritsky, D.  1995, RMexAA,  3,149.

\reference Skillman, E.D, Bothun, G.D., Murray, M.A.  \& Warmels, R.H. 1987, A\&A,  185, 61.

\reference  Tully, R. B.  \& Shaya, E. 1984, ApJ, 281, 31.

\reference van Zee, L.  \& Haynes, M.P. 1998, private communication.

%\reference Van Zee, L., Haynes, M. \& Salzer, J.J. 1997 AJ, 114, 2479
%\reference Van Zee, L. 1996 PhD thesis

\end{references}
\end{document}